\documentstyle[aps,prd,preprint]{revtex}

\tighten
\begin{document}
\draft

\preprint{\vbox{\baselineskip=12pt
\rightline{gr-qc/9412068}
\rightline{Submitted to Physical Review D}}}

\title{Event Horizons in Numerical Relativity I: Methods and Tests}

\author{Joseph Libson${}^{1,3}$, Joan Mass\'o${}^{1,2}$, Edward
Seidel${}^{1,3}$, Wai-Mo Suen${}^{4}$, Paul Walker${}^{1,3}$}

\address{
$^{1}$ National Center for Supercomputing Applications, \\
Beckman Institute, 405 N. Mathews Ave., Urbana, IL, 61801 \\
$^{2}$ Departament de F\'{\i}sica, Universitat de les Illes Balears,
   E-07071 Palma de Mallorca, Spain. \\
$^{3}$ Department of Physics,
University of Illinois, Urbana, IL 61801 \\
$^{4}$ McDonnell Center for the Space Sciences, Department of Physics, \\
Washington University, St. Louis, Missouri, 63130
}
\date{\today}

\maketitle

\begin{abstract}
This is the first paper in a series on event horizons in numerical
relativity.  In this paper we present methods for obtaining the
location of an event horizon in a numerically generated spacetime.
The location of an event horizon is determined based on two key ideas:
(1) integrating backward in time, and (2) integrating the whole
horizon surface.  The accuracy and efficiency of the methods are
examined with various sample spacetimes, including both analytic
(Schwarzschild and Kerr) and numerically generated black holes.  The
numerically evolved spacetimes contain highly distorted black holes,
rotating black holes, and colliding black holes.  In all cases
studied, our methods can find event horizons to within a very small
fraction of a grid zone.
\end{abstract}


\narrowtext

\section{Introduction}

Black holes are among the most fascinating predictions in the theory
of General Relativity.  During the past 20 years there have been
intense research efforts on black holes and their effect on the
astrophysical environment.  With the exciting possibility of detecting
gravitational wave signals from black holes by the gravitational wave
observatories under construction(LIGO and VIRGO~\cite{LIGO3}), we have
seen an increasing surge of interest.  The black hole events
likely to be observed by the gravitational wave observatories involve
highly dynamical black holes, e.g., two black holes in collision.  The
most powerful tool in studying such highly dynamical and intrinsically
non-linear events is probably numerical treatment.  In recent years,
there has been significant progress in numerical relativity in this
direction (See, e.g., Refs.~\cite{Seidel94e,Bona93}).  In
particular, long evolutions of highly dynamical black hole spacetimes
are now possible~\cite{Abrahams92a,Anninos93b}, opening up the
opportunity of many interesting studies.

The defining character of a black hole is its event horizon (EH).  The
EH is defined as the boundary of the causal past of the future null
infinity (for a rigorous description, see, e.g. Ref.~\cite{Hawking73a}).
Photons emitted inside this boundary surface cannot escape to infinity
while those emitted outside in a suitable direction can.  It is the
existence of such a boundary surface that makes a black hole
``black''.  Only when a horizon surface is located can we know for
sure that the spacetime that we are studying contains a black hole.
Locating such horizon surfaces is the subject of this first paper in
our series studying event horizons in numerical relativity.

The study of the dynamics of this surface is crucial for understanding
the dynamics of a black hole spacetime.  In the membrane paradigm of
black holes~\cite{Thorne86}, black holes are characterized by the
properties of its EH, which is regarded as a 2D membrane living in a
3D space, evolving in time and endowed with many everyday physical
properties like viscosity, conductivity, entropy, etc.  We believe
this point of view is powerful in providing insight into the numerical
studies of black holes.  In the second paper in this
series~\cite{Masso95a}, we present methods and tools for studying
these properties of the horizon surface.  In the third paper in the
series~\cite{Masso95b}, we turn to the physics of black holes that can
be explored using these tools.

There are two properties of the EH that make its location
difficult to determine in numerical relativity: It is defined
non-locally both in space and in time, and moreover, in an acausal
manner.  Therefore its location, and even its existence at any time
cannot be determined without knowledge of the complete four geometry
of the spacetime.  In this paper we will present methods that overcome
these difficulties of finding EH's in black hole spacetimes.

Consider a numerically constructed spacetime represented by a set of
data given on a 3+1 lattice.  Suppose the numerical evolution has
covered a region of spacetime so that at the end of the evolution, one
can locate a spatial domain which is likely to contain the EH.  (We
show below how this can be done.)  On the one hand, as numerical
evolution often does not cover null infinity~\cite{Gomez94a},
generally one cannot give a precise location of the EH even on the
final time slice.  On the other hand, the black hole events we are
interested in often involve black holes eventually settling down to a
stationary state after going through violent processes, e.g., the
coalescence of two black holes.  If the numerical evolution covers
late enough times, so that at the end of the evolution the geometry
returns approximately to stationarity, it is often easy to have an
{\it approximate} location of the EH, as we shall show below.  We note
that at late times, if the spacetime has truly returned to
stationarity, the difficulties of finding the EH due to its acausal
nature and non-locality in time disappear, as the geometry is the same
on each time slice henceforth.

Of course, true stationarity is probably impossible in numerical
evolution. One can aim at most at approximate stationarity.
 We note that such approximate stationarity is
possible, as dynamical black hole spacetimes can now be evolved for a
long time (commonly $t\approx 100M$, and in some cases to $1000M$)
\cite{Bernstein93b,Brandt94a,Anninos94e}.
 Therefore the first set of questions in the numerical study of black
hole EH's is this.  Consider a numerically constructed spacetime, such
that one has an approximately stationary final configuration. Assume
that although the EH cannot be located exactly, it is known to exist
in a certain spatial domain. Can one determine this horizon-containing
domain in all the previous time slices during which the spacetime is
highly dynamical, which is often the real epoch of interest?  That is,
how does this horizon-containing domain evolve in a dynamical
spacetime?  We will show that in the language used here, even if the
horizon-containing domain is fairly large at the final time, it can be
narrowed down tremendously at earlier times.  If the EH can be
approximately located at late times, after the black hole and
surrounding spacetime return to approximate stationarity, our result
implies that the EH can be accurately determined throughout the
dynamical period of interest.

The present paper focuses on this set of problems of locating the EH
in a black hole spacetime.  We have developed two generations of
``event horizon finders'', based on two basic ideas~\cite{Anninos94f}:
(1) Integrate backward in time, and (2) Integrate the horizon surface
instead of the individual geodesics.  We discuss these ideas in detail
in Sec.~\ref{sec:ideas}.  In Sec.~\ref{sec:results} we examine how
much accuracy one can obtain in locating the EH throughout its history
(i.e., how small the horizon-containing domain can become) in various
sample numerically constructed spacetimes, including distorted
Schwarzschild and Kerr black holes, and colliding black holes.  We
also compare various methods for locating the EH, including our two
generations of EH finders based on backward integration and another
method recently proposed using forward integration~\cite{Hughes94a}.

A basic assumption in this paper is that one can locate a certain
horizon-containing domain at the end of the numerical evolution.  An
interesting question is to what extent one can use the techniques
developed in this paper to improve the confidence that the
domain truly contains an EH, or to rule out such a possibility.
Although we shall not deal with this question in this paper, it is
easy to see that the techniques developed can also be useful in this
regard.  In Sec.~\ref{sec:results} we shall see that one can easily
narrow this domain on the final configuration to a large extent by
examining trajectories of null surfaces.

\section{The Methods}
\label{sec:ideas}
Suppose for the moment the actual event horizon (EH) is given on the
final time slice of the numerical evolution (for simplicity of
discussion, the evolution is assumed to be in the 3+1 slicing of
spacetime; other schemes, such as 2+2, could be treated as well).  How
does one determine its location at earlier times?  One may consider
integrating null geodesics forward in time, selecting those that
successfully arrive at the given EH at the final time.  There are two
conditions that must be satisfied for success: (i) the geodesic must
land on the given EH, and (ii) the tangent of the geodesics must be
the outgoing normal to the EH.  In principle, such stringent
requirements could be satisfied and the history of the EH could be
traced out with such an algorithm.  However, there is an intrinsic
difficulty making such an algorithm very inefficient.  By the
definition of the EH, outgoing null geodesics slightly inside the EH will fall
into the singularity after some short time, while outgoing null
geodesics just slightly outside will escape to null infinity.
Therefore, evolving null geodesics forward in time near the EH is a
physically unstable process.

In Fig.~\ref{fig:edfink} the behavior of various outgoing null geodesics in a
spherical black hole spacetime is shown in terms of
Eddington-Finkelstein coordinates~\cite{Misner73}.  The horizon is
given as a solid line while photons are given as dotted lines.
Outgoing null
geodesics starting out infinitesimally close to the horizon spread out
to cover a large region of the spacetime.  They are ``everywhere'' on
a late time slice.  This physical instability, coupled
with the inevitable finite differencing error in a numerical
calculation, ensures that null geodesics will move away from the true
position of the EH, even if they are right on it initially.  An
accurate determination of the history of the EH will not be possible
unless (a) the spacetime data has a very high resolution, and (b) a
very large number of photons are followed, so that they are ``dense''
enough to have some of them stay near the EH throughout the evolution.

However, a moment's reflection tells us that this important property
of the EH can be used to our advantage; the instability of integrating
outgoing null geodesics forward in time near the EH implies that when
integrating these geodesics backward in time near the EH, they will
converge onto the horizon.  The backward direction is the
stable direction of integration.  This is one of the two basic
ingredients in building our EH finders.  In our first generator EH
finder~\cite{Seidel93a,Libson93a}, called the ``backward photon''
method, we trace the EH by integrating null geodesics backward in
time.  The geodesic equation is a second order differential equation,

\begin{equation}
{{d^2 x^\alpha}\over{d\lambda^2}} = -{\Gamma^\alpha}_{\beta\gamma}
\frac{dx^{\beta}}{d\lambda} \frac{dx^{\gamma}}{d\lambda},
\label{geodesic}
\end{equation}
where $\lambda$ is an affine parameter along the geodesic.

We note that this backward integration method requires that one first
complete the spacetime evolution and store the entire spacetime
evolution metric, or at least a strip of it which safely contains the
full history of the EH.  However, in all cases we studied, being able
to integrate in the stable direction is well worth the extra storage
requirement.  As the null geodesics are ``attracted'' to the EH when
integrated backward, we found that the position of the EH can be
determined to an accuracy of a small fraction of a grid zone, as shown
in the next section.  Forward integration schemes generally also
require storing the entire spacetime evolution, as many photons must
be ``shot'' forward and then tested to see if they end up inside or
outside the horizon at the final time~\cite{Hughes94a}.

Two comments regarding the ``attraction'' of the EH in backward
integration are in order.  Firstly, this ``attraction'' is only in the
global sense.  Locally, the EH has no special property.  That is,
geometrically the EH is not distinct from other surfaces.  As the
geodesic deviation equation is governed by the Riemann curvature
tensor, which has no special value at the horizon, the outgoing null
geodesics are not attracted to the EH in any local sense.  On the
other hand, in a 3+1 numerical evolution, only those outgoing null
geodesics that are either very near or inside the EH at early times
can remain in the finite range of the spacetime with a non vanishing
lapse which is covered by the numerical evolution at late enough times.  In
other words, when integrating backward in time, outgoing photons in
this range approach the EH after a sufficiently long integration.  The
EH is attractive only in this global sense in time.  Note that this attractive
property implies that our starting point in the backward integration
need {\em not} be exactly on the EH.  This is a key point that we will
quantify below.

Secondly, this attractive property of the horizon is just for outgoing
photons.  An ingoing photon when traced backward will not only leave
the EH surface, it may even leave the finite region of the spacetime.  In
the one dimensional case, inward and outward photons are clearly
distinct (only outgoing ones are displayed in Fig.~\ref{fig:edfink}),
but for the general 3D case, when the two tangential directions of the
EH are also considered, the situation becomes more complicated.  (Here
normal and tangential are meant in the 3D spatial, not spacetime,
sense.)  Whether a
trajectory can eventually be
``attracted'' to the EH or not, and how long it takes for it to become
``attracted'', depends on the photon's starting direction of motion.
We note that even for a photon which is already exactly on the EH at a
certain instant, if its velocity at that point has some component
tangential to the EH surface (as generated by, say, numerical
inaccuracy in integration), the photon will move outside of the EH
when traced backward in time.  For a small tangential velocity, the
photon will eventually return to the EH.  The duration and distance it
moves outside the EH depends on its tangential velocity.  Also, the
position to which it returns will not be the original position.

This kind of tangential drifting is undesirable not just because
it introduces inaccuracy in the location of the EH, but more
importantly, because it can lead to spurious dynamics of the ``EH'' thus
found.  Neighboring generators may cross, leading to
numerically artificial caustic points (for an introductory discussion of
caustics, see, e.g., Ref.~\cite{Misner73}, Chapter 10.)

Unfortunately such tangential drifting is not easy to avoid due to the
nature of the geodesic equation (1) which we integrate backward.
The geodesic equation is not only second order, but also requires
derivatives of the numerically generated metric data which are
generally more inaccurate.

Another consequence of the second order nature of the geodesic
equation is that not just the positions but also the directions must
be specified in starting the backward integration.  Neighboring
photons must have their starting direction well correlated in order to
avoid tangential drifting across one another.  We find it important to
make explicit use of the property that the surface we wish to trace is
a closed 2D surface embedded in the 3D space at the start of the
backward integration.  While the starting positions of the photons are
taken to be on this surface, the starting directions are taken to be
normals to this surface.  Of course even if the geodesics have
accurate starting values they may still drift due to inaccuracy in
integration, as discussed above.  We find that the ``backward photon''
method is still quite demanding in finding an accurate history of the
EH, although the difficulties are much milder than those arising from
the instability of integrating forward in time.

In our second generation horizon finder, building on the above idea of
explicitly using the surface forming property of the EH, we follow the
entire horizon surface itself, rather than tracking individual photons
independently.  Generally speaking, the EH can be considered as a 2+1
null surface, except at special points where its normal cannot be
defined.  Except for these special points, which we shall discuss
later, one can represent the 2+1 EH surface by a function
\begin{equation}
f(t,x^i) = 0,
\label{ff}
\end{equation}
which
satisfies the null condition
\begin{equation}
g^{\mu\nu} \partial_{\mu} f \partial_{\nu} f = 0.
\label{nullsurface}
\end{equation}
Hence the evolution of the surface can be obtained by a simple
integration,
\begin{equation}
\partial_t f = \frac{ - g^{ti} \partial_i f +
\sqrt{(g^{ti}\partial_i f)^2 - g^{tt} g^{ij} \partial_i f \partial_j f}
}{g^{tt}}.
\label{evolve}
\end{equation}

In our second generation horizon finder, we integrate this equation
backwards in time.  This is what we call our ``backward surface''
method.  Notice that at the final time, the EH is given as a closed 2D
surface.  Given this as the starting condition for the backward
integration of Eq.~(\ref{evolve}), there is no other condition needed
(e.g., there is no need to specify either an initial direction  or
boundary conditions for the surface).  The reconstruction of the
complete function $f(t,x^i)$ gives us the full history of the EH (in
fact, much more than this, as we will discuss below and detail
in Ref.~\cite{Masso95a}).

Both the backward photon and backward surface method work very well
as shown in Sec.~\ref{sec:numerics} below.
However, there are a number of advantages of using the
backward surface method:

\noindent {\it (i)} Of primary importance is that the method is
simpler and less susceptible to numerical error than the backward photon
method.
Notice that Eq.~(\ref{evolve}) contains only derivatives of the
surface and {\em not} of the metric components themselves and is
therefore less susceptible to the numerical inaccuracies present in
the metric data.  The horizon is generally found in numerically
evolved spacetimes in regions where metric functions contain rather
steep gradients that are poorly resolved~\cite{Seidel92a}, and hence
their derivatives may not be accurately known there.

\noindent ({\it ii}) Tangential drifting is not a source of error,
because the only direction that a surface can move is normal to
itself.  Once the surface becomes the EH, it cannot drift away from
it.  This is in contrast to integrating geodesics; even if the
geodesic is right on the horizon at some instant in time, it does not
guarantee that it can stay on it all the time.  See
Fig.~\ref{fig:tandrift} below.

\noindent ({\it iii}) Unlike integrating null geodesics, the result of
this method is guaranteed to be surface forming, as we are explicitly
integrating surfaces.  This is a nontrivial advantage over integrating
geodesics; in particular, the integrated geodesics can fail to be
surface forming, either due to the numerical error discussed above, or
due to the existence of caustics on the horizon.  At the caustics the
geodesics leave the EH surface when integrated backward in time.  If
there are geodesics leaving the EH through caustics, but this caustic
point is not so recognized, and the EH is taken to be the surface
connecting all geodesics, the evolution history of it would be
completely wrong.  However, it is nontrivial to determine
whether a caustic or numerical error causes  geodesics to
become non-surface forming.  In both cases, we may see null geodesics
cross one another and move outward.  See Fig.~\ref{fig:tandrift}
below.

\noindent ({\it iv}) The surface method is naturally suitable for
handling and studying caustic structures on the EH.  (Here for
simplicity of discussion we indiscriminately refer to all classes of
points at which horizon generators leave the EH when traced backward
in time as horizon caustics, see, e.g., Ref.~\cite{Misner73}.)  As the
normals of the EH are different when the caustic point is approached
in different directions, strictly speaking the EH is not a null
surface at such points, and Eq.~(\ref{evolve}) would seem to be in
difficulty.  However, we note that in the generic case of an isolated
caustic point, the EH surface surrounding the caustic point can be
evolved using one-sided spatial derivatives on the RHS of
Eq.~(\ref{evolve}).  Despite the fact that there is no well defined
normal to the surface at a caustic, the motions of surface elements on
different sides of the caustic must be continuous as determined by
Eq.~(\ref{evolve}) with appropriate one sided derivatives, as the EH
is a continuous closed surface at each time slice.  In the two black
hole collision case studied below in Sec.~\ref{sec:twobh},
Fig.~\ref{fig:twobhphot}, there is clearly a cusp in the EH along the
$z$-axis, so that the surface is not smooth there, but it is continuous.
This is obtained with a one-sided treatment of the derivative there.
This treatment requires the knowledge of where caustics would be
forming {\em a priori}.  Such {\em a priori} knowledge is often
possible for spacetimes with symmetry properties.  In fact, all
dynamic black hole spacetimes evolved to date (of which we are aware)
have such properties.  At those points where caustics may appear,
suitable one-side derivatives are used.

Alternatively, one can treat horizon caustics with the surface method
{\em without} introducing special treatment for possible caustic
points, provided that we do not restrict Eq.~(\ref{evolve}) to
describe {\em only} the EH, but {\em also} the locus of the null
generators which leave the EH through the caustic points backward in
time.  That is, in tracing Eq.~(\ref{evolve}) backward, we allow the
closed surface to cross itself.  The point of crossing is the caustic.
Although the normal of the surface is not continuous going across the
caustic on the EH, it is continuous going from the EH across the
caustic point to the locus of generators which has left the
EH~\cite{Thorne94b}.  This is true provided a suitable identification
of neighboring surface elements is used, namely, when the
identification of two neighboring surface elements does not change in
time.  This issue is considered in this paper where we treat the
collision of two black holes in section \ref{sec:twobh}, when caustics
are important.

These backward surface and photon methods bring for the first time the
possibility of studying the properties of horizon generators and caustics in
numerical relativity, but the surface method provides a particularly
elegant, economical, and accurate way of computing this structure.

Before we go on to the next subsections on the tests and the accuracy
of the methods, we comment on the fact that the starting position of
the EH in the backward integration is often not known precisely.  As
pointed out earlier, as long as the numerical evolution can be carried
to a point that the black hole returns to approximate stationarity, it
is often possible to locate a region which contains the EH.  For
example, the apparent horizon (AH) is always inside the event horizon
(provided quantum effects are ignored).  If an AH is found, it can be
taken to be the inner boundary of this horizon-containing region.  The
real task in locating the horizon is to determine the evolution of
this horizon-containing region.  Due to the attractive nature of the
null surfaces to the EH, the horizon-containing region can be narrowed
substantially in the backward integration.  Indeed we shall see that,
using our methods, it is often easy to narrow this region down to much
less than a grid separation used in the numerical construction of the
spacetime.  In some sense the location of the EH is determined to a
precision higher than the resolution of the background spacetime,
something seemingly impossible at first sight.  This is not paradoxical
as in all cases studied in this paper, the EH surface expressed as a
function, Eq.~(\ref{ff}), is constructed using information and
interpolations involving many data points, hence ``washing out'' some
local fluctuations.  However, if the horizon-containing region were to just
span a few angular grid points, the localization of it to a small
fraction of a grid separation would no longer be meaningful.

Using our backward methods we are able to trace accurately the entire
history of the EH, as we detail in the next section.  However, there
are cases in which in the region where the numerically constructed
spacetimes are badly resolved (e.g., the crotch region in the two
black hole case study below), where the backward surface method method
is capable of producing more reliable results.  The basic difference
in the two methods is in their computational requirements and
convenience. A typical case studied here is that of a black hole
interacting with a gravitational wave.  Such a case is resolved on a
grid of 200 radial by 53 angular zones, and evolved to $t=75M$.  To
trace the EH to 1/10 of a grid separation for the dynamical period of
the evolution ($0M < t < 48M$) takes only a few minutes on a computer
workstation.  For the backward photon method to achieve the same
accuracy, it takes several times longer.  For future applications with
dynamical black hole spacetimes evolved to thousands of $M$, we
believe the backward surface method is most promising.

\section{Numerical Techniques and Tests of Methods}
\label{sec:results}
In this section we discuss both the numerical implementations of our
surface methods and provide examples of their applications to dynamical
spherical black holes, distorted axisymmetric black holes with and
without rotation, and colliding black holes.

\subsection{Numerical Implementations}
\label{sec:numerics}

First we discuss details of the numerical implementation of these
methods.  For the backward photon method, we use a standard adaptive
step size, fourth order in time Runge-Kutta method~\cite{Press86} to
integrate the second order geodesic equation~(\ref{geodesic}).  For the
surface method we have used a number of methods, including second
order leapfrog, a second order MacCormack predictor corrector method,
and a fourth order Runge-Kutta method using the method of lines to
integrate hyperbolic Eq.~(\ref{evolve}) in time as a set of coupled
ordinary differential equations.  All methods give similar results,
but for highest accuracy, we use the fourth order Runge-Kutta method
of lines to generate the data presented here.

In following the horizon backward through the spacetime, we
necessarily require the spacetime data at points that do not lie on
the numerical grid.  For this we must interpolate the spacetime data
to the actual location of the horizon at each time.  For the backward
photon method, we must interpolate {\em both} the metric and its
derivatives to these locations.  On the other hand, for the surface
method we need {\em only} the metric itself.  For both methods, we
find that second order interpolation is adequate to determine these
data values.

In the surface method, the location of the surface is represented by a
function.  The use of a suitable parameterization of the surface
is important.  For the axisymmetric cases discussed in this paper, a
convenient choice is
\begin{equation}
f(t,r,\theta) = r - s(\theta,t) = 0.
\label{surfaceparm}
\end{equation}
With this parameterization, Eq.~(\ref{evolve}) for the evolution of
the surface becomes
\begin{equation}
\partial_t s = -\frac{ - g^{tr} + g^{t\theta} \partial_\theta s +
\sqrt{(g^{tr} - g^{t\theta} \partial_\theta s)^2 - g^{tt} (
g^{rr} - 2g^{r\theta}\partial_\theta s +
g^{\theta\theta} (\partial_\theta s)^2)}
}{g^{tt}}.
\label{surfevolve}
\end{equation}
The angular derivatives of the function $s(t,\theta)$ are computed
using both second and fourth order finite difference methods, with
similar results.  Although this way of representing the surface works
well for almost all cases discussed in this paper, as we show in
section~\ref{sec:twobh}, other parameterizations can be necessary at
times.

\subsection{Test Beds and Examples}
\subsubsection{Spherical Black Holes}
\label{sec:onebh}

In this section we show how the methods detailed above can be applied
to various different single black hole spacetimes.  The first case we
consider is a pure, spherical Schwarzschild black hole, evolved with
the 2D, axisymmetric black hole code described in
Refs.~\cite{Anninos93c,Bernstein93b}.  We use this important testbed
case to show the accuracy to which one can determine the location of
the horizon in a numerically evolved spacetime.  Although the
spacetime is geometrically the static Schwarzschild spacetime, it is
evolved numerically with a maximal slicing condition which makes the
metric functions change in time.  As discussed in
Ref.~\cite{Seidel92a}, such time dependence makes even Schwarzschild
quite difficult to evolve numerically for long periods of time (beyond
about $t=100M$ with reasonable grid parameters).  As the coordinates
fall in towards the hole the horizon moves out in coordinate space,
the lapse collapses, and large gradients develop in the metric
function near the horizon.  (For more
details on these problems, see Refs.~\cite{Seidel92a,Bernstein93b}.)
The advantage of using this as our first testbed case is that, on the
one hand, the numerical spacetime constructed with this code has many
of the properties and difficulties of a general numerically
constructed black hole spacetime.  On the other hand, the spacetime is
really a Schwarzschild spacetime for which we know where the EH should
be for all time.  In particular, in this case the apparent horizon
(AH) and EH coincide.  We have accurate AH finders~\cite{Anninos93a}
that can locate the AH, and thus in this case the EH; on any given
single slice of the spacetime we know both horizons without needing to
know the future or past of that slice.  This provides us with
important accuracy checks on our methods of locating the EH throughout
the evolution.

In Figs.~\ref{fig:ehahsph}a-d we show results for a spherical black hole
spacetime evolved with our axisymmetric 2D black hole code.  We apply
our horizon finder to the data obtained in the evolution assuming
neither spherical symmetry, nor the fact that the spacetime geometry
is really Schwarzschild.  At the final time slice, the
horizon-containing region is determined by examining the lapse
function and the radial metric function.  For a spacetime evolved with
maximal slicing, the event horizon resides in a region
with a partially collapsed lapse function.  In
Fig.~\ref{fig:ehahsph}a, we show the lapse function, $\alpha$, at the
final time slice, $t=100M$.  We take the horizon-containing region to
extend from $\alpha=0.1$ to $\alpha=0.7$, with {\em (o)} labeling the
outer edge and {\em (i)} the inner edge of it.  In
Fig.~\ref{fig:ehahsph}b, the radial coordinate $r$ of the two
surfaces {\em (o)} and {\em (i)} traced backwards in time is shown.
At $t=100M$, the two surfaces are separated
in the radial coordinate by $3.4M$.  By $t=85M$, the two
lines are separated by just one grid zone, corresponding to a
difference in $r$ of $0.35M$.  By $t=70M$, the two lines are no longer
distinguishable, with a separation down to $1/10$th of a grid zone,  a
difference in $r$ of $0.03M$.
The separation exponentially decreases down to $1\times10^{-6}$ grid zones
at $t=0M$.  This rapid shrinking of the horizon-containing region is a
direct consequence of the divergence of null geodesics forward in time
shown in Fig.~\ref{fig:edfink}. We conclude that if the aim is to locate
the horizon to one grid zone accuracy, we have succeeded in doing so
for the times $t=0M$ to $t=80M$.  We emphasize that no information
about the apparent horizon is used in the process.

For the purpose of comparison, in Fig.~\ref{fig:ehahsph}b we have also
shown the trajectory of surfaces extremely far outside and far inside the
horizon-containing region.  These surfaces are shown as dashed lines.
We see that the outer one converges quickly to the other test
surfaces, while the inner one is initially trapped in a region of
collapsed lapse.  At $t=40M$, all the surfaces are practically
indistinguishable.

In Fig.~\ref{fig:ehahsph}c, we show the evolution of the coordinate
locations of these surfaces in the first quadrant.  The surfaces marked
{\em (i)}, {\em (o)} and AH are the same surfaces as shown in
Fig.~\ref{fig:ehahsph}b.  Here, we have evolved an additional,
nonspherical, surface.  The location of this surface is given at
$t=100M$ by the formula
\begin{equation}
\eta=\eta_0 + A \cos w\theta,
\label{distort}
\end{equation}
with $\eta_0$ chosen to be the radial position of the apparent
horizon, with $w=4$ and $A=0.2$.  We evolve this surface to
demonstrate that our initial trial surfaces need not have the same
angular dependence as the EH (in this case, spherical).  In a general
dynamical black hole spacetime, it will not be possible to pick trial
surfaces having the same coordinate or geometrical angular dependence
as the EH to be traced out.  Such trial surfaces are not necessary,
though.  In the case shown in Fig.~\ref{fig:ehahsph}c, where the trial
surface is quite non-spherical, with part of the surface inside and
part outside of the EH, we see that the trial surface quickly
converges when traced backwards in time.  All of the surfaces are very
close and almost completely spherical by $t=70M$.  By $t=50M$, all the
surfaces are within $1/10$th of a grid zone.  We note, however, that a
sufficiently nonspherical surface may itself develop caustics,
particularly if it is initially far from the true EH. In
Fig.~\ref{fig:ehahsph}d we show the evolution (at times $t=98.5M$,
$t=98.4M$, $t=98.3M$ and $t=98.2M$ from top to bottom, with the line
marked AH labeling the position of the AH and thus true EH at
$t=98.5M$) of a highly distorted surface with the angular dependence
of Eq.~\ref{distort} increased to $w=16$, the amplitude decreased to
$A=0.1$, and the center of the perturbation moved away from the
apparent horizon. We find that the surface method fails, with
numerical noise developing as the surface tries to cross itself.  This
crossing in itself is not fatal to the surface method, but the
particular parameterization (\ref{surfaceparm}) of the surface cannot
describe this crossing.  As we see below, the self crossing of the
surface can be handled with a proper parameterization, which is needed
when {\em true} caustics develop, as in the collision of two black
holes.  We stress that when the black hole has returned to
quasi-stationarity at late times, one does not expect the EH to have
such rapidly varying angular dependence and one would not pick such
surfaces as the outer or inner boundaries of the horizon-containing
region. We study such an extreme, contrived example only to explore
the limits of our methods.

We note that at late times the data representing the spacetime itself
as obtained in Refs.~\cite{Seidel92a,Bernstein93b} becomes inaccurate due
to the large spikes developing in the metric functions near the
horizon~\cite{Seidel92a}.  As these spikes become ever steeper during
the evolution, they become less and less well resolved, and therefore they
are not accurately modeled on the numerical grid.  This lack of
accuracy in the spacetime itself is reflected in the calculation of
the area of the apparent horizon at late times, which increases rather
than remaining constant.  Remarkably, this lack of accuracy at
late times, when the EH finding algorithm is started, does {\em not}
cause any difficulty in finding the EH at earlier times, as seen in
the figures.  Therefore, not only are our algorithms able to find
accurately the true EH even with a poor initial guess for its
location, they are also insensitive to inaccuracies in the
spacetime data that inevitably occur at late times.

Finally, in Fig.~\ref{fig:tandrift}, we show the
tangential drifting that can occur with the backward photon method,
as discussed in Sec.~~\ref{sec:ideas}.  Fig.~\ref{fig:tandrift} also serves as
an illustration
to the other comment we made above
concerning the ``attractiveness'' of
the EH to backward integrated photons, namely, the attraction is only in
the global sense.
In this example the tangential drifting is
due to the choice of the initial direction of integration.
Two photons are traced backward
(shown as dotted
lines) beginning at the exact location of the EH at $t=100M$, but
with a
3\% error in the starting direction.  That is, instead of being
normal to the EH
($ p_ \theta/ p_r =  0 $, in obvious notations),
we use $ p_\theta / p_r  =  \pm0.03$.
In Fig.~\ref{fig:tandrift}, the trajectories of these photons are shown
with the corresponding times marked.  The radial coordinate are
normalized by the position of the EH, so that the EH is at $1$ on the vertical
(radial coordinate $\eta$) axis.  We see that with the 3\% error in the
starting directions of the photons, the photons move out of the EH when traced
backward in time.   If these photons were taken as horizon generators,
this would introduce a small error (note the scale of the
$\eta$ axis) in the location of the ``EH'' for a period of
time, as the photons are gradually ``attracted'' back to the correct radial
location after some integration.  However, the error in the tangential
direction is substantial, as we can see in Fig.~\ref{fig:tandrift},
where the horizontal axis is given in terms of $\theta$ in radians.
In particular, the two photons cross each other, creating an
artificial ``caustic'' at $t=98.0M$ on their way out from the EH.
Although they return to the correct radial EH location eventually, their
$\theta$
values change dramatically, making the trajectories very
different from those of the true horizon generators.

The case just discussed was a vacuum Schwarzschild black hole with no
true physical dynamics, although the motion of the coordinates through
the spacetime makes both the black hole evolution and tracking of the
EH nontrivial.  In order to break the degeneracy between the AH and
EH, we next consider a non-vacuum case.  In this case we
evolve a spherical black hole with a relativistic, massless
Klein-Gordon scalar field falling into it.  The system is described by
the Klein-Gordon equation,
\begin{equation}
g^{\mu \nu} \phi_{; \mu ; \nu} = 0,
\label{scalar}
\end{equation}
coupled to the Einstein equation through the energy momentum tensor of
the scalar field.  This problem has been studied previously in
Ref.~\cite{Seidel92a}.

With a gravitating scalar field falling into the black hole, the
system has true physical dynamics.  Not only does the horizon move out
as coordinates fall into the hole, as above, but now the horizon also
expands in a geometric sense, as its area must increase to accommodate
the infalling matter.  A particular test case is shown in
Fig.~\ref{fig:scalar}.  Here the scalar field was set up in a gaussian
shell surrounding the black hole.  As the field propagates into the
hole, the horizon expands as expected.  The solid line labeled
``Event Horizon'' is obtained by integrating the surface backward
in time from the AH location at time $t=60M$.  We refer to this line
as the ``EH'' because as shown above it will converge rapidly to the
true EH .  The dashed line is the AH obtained by
solving the AH equation~\cite{Anninos93a} at each time slice.  Note
that the AH always lies inside the EH, as expected, and that at early
times the EH is quite a bit larger as it starts expanding before the
incoming matter arrives.  The two solid lines labeled ``Escaping
Photon'' and ``Trapped Photon'' were obtained by integrating radially
outgoing null geodesics {\it forward} in time from locations 1/10th of
a grid zone inside and 4 grid zones outside the EH at early times.

We note that even in this spherically symmetric case it is not
possible to integrate accurately the path of the event horizon {\it
forward} in time.  Even when photons were placed right on the known
horizon position initially, due to the unstable nature of the forward
integration, the horizon could be tracked for only a short while
before it would diverge away from the true EH.  If we want to track
the horizon further in time, one way is to ``abandon'' the original
photons, and start new ones closer in to the horizon at the time when the
original ones get too far apart, but the resulting horizon will
consist of a discontinuous surface. A variation on this idea would be
to use forward photons simply as a probe. If any photon integrated
forward in time from a given point is not within the apparent horizon
at late times, then that point may be considered outside the EH.  By
integrating many photons forward from many spacetime points, a horizon
surface can be mapped out, as shown in Ref.~\cite{Hughes94a}. This is
an effective but time consuming procedure, and the trajectories traced
do not give the trajectories of the the generators of the horizon.

\subsubsection{Distorted Axisymmetric Black Holes}
\label{sec:distorted2dbh}

Next we present results for an axisymmetric black hole that has been
distorted by the presence of a gravitational wave.  The initial data
sets for these studies consist of an Einstein-Rosen bridge in the
presence of a gravitational wave in the form originally considered by
Brill, and have been described extensively in
Ref.~\cite{Bernstein94a}.  The gravitational wave is set up as a torus
surrounding the black hole, and its location, amplitude, and shape can
be varied essentially arbitrarily.  The code used to evolve these data
sets numerically has been discussed in
Refs.~\cite{Abrahams92a,Anninos93c,Bernstein93b}.

The first case we study here is that of a black hole with a narrow
ring of weak gravitational waves isolated from the hole initially.  When
this system is evolved, we expect the hole to become distorted
slightly as the waves impinge on it, and then settle down to a
Schwarzschild hole with a larger mass afterwards.  Such a case
provides a strong test of our methods as we must be able to track a
horizon that begins essentially as a sphere, develops a distortion as
it is hit by a nonspherical wave, and returns to a sphere after a long
integration.

In Fig.~\ref{fig:pertarea}a, we demonstrate that we can locate the EH
to high accuracy for such a spacetime.  At time $t=75M$, the black
hole has returned to approximately the Schwarzschild geometry.  The
apparent horizon, represented as the short dashed line in
Fig.~\ref{fig:pertarea}a, has returned to an almost exact sphere.  As the EH
must lie outside the apparent horizon, the inner boundary of the
horizon-containing region can be taken to be the location of the
apparent horizon on the last time slice, $t=75M$.  The outer boundary
is taken to be the line marked {\em (o)} in Fig.~\ref{fig:pertarea}a,
representing a spherical surface some distance outside the apparent
horizon.  The fact that this line is safely the outer boundary can be
seen, as the area of this spherical surface is shrinking backward in
time.  On the other hand, the line marked {\em (i)} starting from the
AH at $t=75M$ is expanding outwards.  The two surfaces exponentially
approach each other, and this separation becomes less than a grid zone
at $t=62M$.  Note that at this time, this one-grid-point-wide
horizon-containing region is entirely outside the AH.

In Fig.~\ref{fig:pertarea}a, we also show other test surfaces as long dashed
lines which are well outside or inside the horizon-containing region.
All these test surfaces coincide to much less than 1/10th of the grid
separation (corresponding to typical proper distances between the
surfaces of less than $0.01M$) for the range $t=0$ to $40 M$. The
insert shows an expanded view of the early time.  All surfaces
computed are shown, but they are completely indistinguishable in spite
of their extremely different starting positions, clearly showing the
power and stability of this method.  For all practical purposes, this
surface can be regarded as the EH.  At $t=0$ the AH and EH
practically coincide with each other.  Then the EH foresees the coming
of the wave and expands.  As the wave is falling in, after about
$t=15M$, the AH starts to expand and catch up. The behavior of the AH
and EH are exactly as expected. (We note that
the area of the hole continues to drift up after 40M.  This is a well
known numerical inaccuracy due to the development of a sharp peak in
the radial metric component. This effect is in the background spacetime
and is unrelated to our schemes for locating the horizon.)

In Fig.~\ref{fig:pertarea}b, we show the maximum separation over the
whole surface between the outer and inner boundary of the
horizon-containing region versus time (maximum among the angular
zones).  The vertical axis is in terms of grid separation.  As marked
by squares, the maximum separation exponentially decreases down to 1,
at $t=60M$, then keeps decreasing to $1/10$, $1/100$, $1/1000$,
$1/10,000$ and $1\times 10^{-5}$ of a grid zone by $t=50, 39, 27, 16$,
and $0M$ respectively. Again, for all practical purposes the outer and
inner boundary of the horizon-containing region coincide, and the
region can be regarded simply as the location of the EH for earlier
times.  We also note that at $t=0$ for the present case, the EH
surface is found to be geometrically spherical (using the tools
described in the second paper in this series) to within 1 part in
$10^6$.

Finally, we also compare the result of our backward photon method to
our surface method in Fig.~\ref{fig:pertarea}c.  We show the
coordinate location of the surfaces integrated directly according to
Eq.~(\ref{evolve}) at various times, and also the location of surfaces
formed by integrating the geodesic equation (Eq.~(\ref{geodesic}))
backward in time.  For the latter integration, the initial locations
of the photons
were taken from the initial trial surface, and their initial
directions were taken to be normal to the surface.  In this figure the
results are indistinguishable, as they coincide to
within 0.05 grid zones throughout the
evolution.  We note that we have also
compared the results of the forward integration technique described in
Ref.~\cite{Hughes94a} to our backward methods. For this forward
integration technique to determine the EH to similar accuracy as we
have done here, however, would take around $10^5$ times more CPU time
than our backwards integration methods.

Next we briefly consider a more strongly distorted non-rotating black
hole, providing a test of our methods in the more nonspherical and
highly dynamic regime.  In this case we choose initial data parameters
($Q_0$,$\eta_0$,$\sigma$,$n$) = (1.0,0,1.5,2) in the language of
Ref.~\cite{Anninos93a}.  This black hole is significantly more
distorted than the previous case (the ratio of the polar to equatorial
circumference of the apparent horizon is 4.26 at the initial time).  In
Fig.~\ref{fig:distortcoord} we show the maximum width of
the horizon-containing region over all angular zones.  The initial
width of the horizon contain region at
$t=60M$,  chosen in the same manner as above, is 10 grid zones.
We see that this highly non-spherical case is no different from the previous
cases.  The width of the horizon containing region
decreases exponentially as a function of time, being less than 1
grid zone at $t=48M$ and being only $0.0001$ at $t=0$.  Again, we mark
the points where the width of the region is $1/10$,$1/100$, and $1/1,000$th of
a zone in the figure.

\subsubsection{Distorted Rotating Black Holes}
\label{sec:kerrbh}
The last single black hole cases we consider are rotating black holes.
Rotating black holes are expected to be the end point of all
astrophysical black hole systems, so they are essential cases to be
considered.  Rotation adds a new dimension to the problem, as new
metric elements are involved, another polarization of the
gravitational wave is present, and horizon generators will now be
dragged around the black hole due to its angular momentum.
Furthermore, the horizon of a Kerr black hole is not spherical, but is
oblate, with the oblateness related to the rotation parameter of the
hole~\cite{Smarr73a,Brandt94b}.  For these reasons the rotating case provides
not only an important testbed, but also a rich area for the study of
horizon dynamics.

First, we consider the Kerr spacetime, which is known analytically.
In Ref.~\cite{Brandt94a}, the Kerr spacetime is studied in a coordinate system
like that used for the studies of distorted Schwarzschild black holes
discussed above.  We have taken the analytic Kerr metric in these
coordinates, with a rotation parameter $a/m=0.68$, to study our EH
finders for rotating spacetimes where the EH location is known
analytically.  In Fig.~\ref{fig:kerr} we show the maximum coordinate
separation between three test surfaces and the analytically known
horizon versus time.  Again, we see an exponential convergence of the
surfaces to the exact location.

Finally we consider a distorted, rotating black hole data set evolved
with a code described in Ref.~\cite{Brandt94a}. In this case the black
hole has been distorted by an axisymmetric gravitational wave, similar
in construction to the distorted Schwarzschild black hole data sets
described above.  This particular data set corresponds to a dynamic
rotating hole that evolves to a Kerr spacetime with a large rotation
parameter of $a/m=0.82$.  In Fig.~\ref{fig:kerrdistort}a, we show the
evolution of the coordinate locations of various test surfaces at
various times.  At $t=60M$, the line marked {\em (o)} represents a
coordinate sphere with an almost constant $\alpha =0.7$.  The surface
marked {\em (i)} is approximately the late time apparent horizon.  The
horizon-containing region is bounded between these two surfaces marked
{\em (o)} and {\em (i)}.  Other test surfaces, shown as dotted and
dashed lines, are also evolved for comparison.  We see that by $t=43M$
these surfaces have converged, and by $t=30M$ all the surfaces are
essentially indistinguishable.  In Fig.~\ref{fig:kerrdistort}b we show
the width of the horizon-containing region [maximum differences in
grid zones between the two trial surfaces, {\em (o)} and {\em (i)}] as
it diminishes exponentially. It is less than one grid separation by
$t=50M$.  Again, $1/10$, $1/100$, and $1/1,000$ of a grid zone
separation are marked on the figure.

We note that although the coordinate shapes of these rotating black
hole horizons are fairly spherical, their intrinsic geometries are
highly distorted.  For an analytic Kerr spacetime, if the rotation
parameter exceeds a critical value of $a/m = 0.867$, the horizon
geometry becomes so distorted that it cannot be embedded into a
Euclidean space~\cite{Smarr73a,Brandt94a}.  When a gravitational wave
is also present in the system, the horizon geometry becomes more
distorted and evolves in time.  Therefore, the cases presented here,
with rotation parameters of $a/m = 0.68$ (Fig.~\ref{fig:kerr}) and
$a/m = 0.82$ (Figs.~\ref{fig:kerrdistort}a-b) have highly nontrivial
horizon geometries, and our horizon finder converges to the true
horizons quickly and tracks them very accurately.  The geometry and
physics of the EH will be analyzed in the third paper in this
series~\cite{Masso95b}.

\subsubsection{Colliding Black Holes}
\label{sec:twobh}
In this section we focus on extracting the event horizon from data
representing the collision of two equal mass black holes.  The
evolution of the spacetime itself has been treated extensively in a
series a papers~\cite{Anninos93b,Anninos94a,Anninos94b}, and we will
not go into the details of those calculations here.  As we discussed
in Sec.~\ref{sec:ideas} above, both the backward surface and photon
methods can be used to study horizons in the collision of two black
holes.  We use these techniques here to find the EH for the two black
hole spacetime parameterized by the quantity $\mu=2.2$, corresponding
to two black holes separated initially by a proper distance of
$L=8.92M$.  We also note that due to the nature of the Cadez
coordinates in which the evolution takes place, we can still use the
parameterization described in Eq.~(\ref{surfaceparm}).  (See
Ref.~\cite{Anninos94b} for definitions of these parameters and the
Cadez coordinate system.)

In Figs.~\ref{fig:twobhcompare}a-c we show results of integrating the
boundary surfaces of the horizon-containing region backward in time,
starting at a late time ($t=75M$).  At this time the two holes have
already coalesced, forming a single larger, almost stationary, black
hole.  As the AH is readily found in this case, we can use it as the
inner boundary of the horizon-containing region.  In
Fig.~\ref{fig:twobhcompare}a, the lapse at $t=75M$ is shown in a 2-D
$\rho-z$ coordinate plot. The surface marked {\em (i)} coincides with
the AH at final times. The lapse on the AH has a nearly constant value
$\alpha=0.34$.  The distribution of the lapse is basically spherical
at this late time, signaling that the geometry is approaching that of
Schwarzschild. (It is basically Schwarzschild except in the innermost
part where the lapse is practically zero. Notice that one of the
throats, where the lapse is exactly zero, can be seen on the $z$
axis.)  Again, the outer boundary of the horizon-containing region,
denoted {\em (o)}, is taken to be a sphere of almost constant $\alpha
= 0.7$.  In Fig.~\ref{fig:twobhcompare}b, the evolution of the
surfaces at various times is shown.  The {\em (i)} and {\em (o)}
surfaces are indistinguishable even at $t=50M$.  To show the
convergent effect more clearly, again we show a highly non-spherical surface
as a short dashed line.  From $t=40M$ backwards, all lines are
indistinguishable.

In Fig.~\ref{fig:twobhcompare}c, the maximum width of the
horizon-containing region between surfaces {\em (i)} and {\em (o)} in
terms of grid separation is plotted versus time, with the points of
$1/10$, $1/100$, $1/1,000$, $1/10,000$, and $1/100,000$ of a grid zone
separation marked.
If the aim is to locate the EH to 1 grid separation, we have achieved
that in the range $t=0M$ to $t=58M$, fully covering the epoch of interesting
dynamics of the coalescence of the two black holes.  (It also
covers more than thrice the dynamical timescale of the final system, as the
natural period of the final black hole is $16.8M$).

As discussed in Sec.~\ref{sec:ideas}, some implementations of the
backward surface method may require special treatment on the axis of
symmetry.  There is a caustic point that must develop where the
horizon surface intersects the $z$-axis.  At this point a cusp
develops in the horizon, causing its normal to become discontinuous.
This means that the surface can hit the $z$-axis at an angle, as one
can see clearly in Fig.~\ref{fig:twobhcompare}b for the surface marked
$t=0$ (i.e., the EH is not perpendicular to the axis on the line between
the holes.)  Because of the symmetry involved in this problem, we know
in advance where this happens, so it is easy to devise numerical
treatment to handle this special situation.  In evolving
Eq.~(\ref{evolve}) one requires derivatives of the surface.  These
derivatives are well defined everywhere except at the caustic point on
the $z$-axis, where only one-sided derivatives are defined.  In
practice we find that in numerical evolution of the surface one can
use a uniform treatment of one-sided derivatives all along the
boundary, including both the equator and the $z$-axis where the cusp
develops.

In Fig.~\ref{fig:twobhphot} we show a comparison of using backwards
geodesic integration and the backwards surface method described above.
It is clear that the backwards surface method and backwards photon
method give the same result, agreeing to within $0.05$ grid
separations throughout the calculation.  As these methods are
completely independent, sharing only the spacetime data to which they
are applied, in each case they confirm that the horizon has been
accurately found, even at the caustic point.

As we discussed in Sec.~\ref{sec:ideas}, by considering the entire set
of null rays generating the horizon surface, including those that have
not yet joined the horizon, one can also treat this problem using the
backward surface method in a way that does not require any special
treatment at the cusp.  We allow the surface pass through itself going
backwards in time where the caustic line forms~\cite{Thorne94b}. In
this way we can trace out the set of generators that will join onto
the horizon in the future.

However, when the locus of generators is included in the evolution, the
parameterization given by Eq.~(\ref{surfaceparm}) is not suitable as it
may become multiply valued.  The evolution equation of the surface
method (Eq.~(\ref{evolve})) gives no restriction on the parameterization.
We choose to integrate the surface in cylindrical $\rho-z$ coordinates
in this case, choosing
\begin{equation}
f(z,\rho,t) = \rho - s(z,t),
\label{fflocus}
\end{equation}
as the surface is single valued in $z$.

In Fig.~\ref{fig:twobhlocus} we show the locus of generators as it is
evolved by the code.  The locus of generators clearly passes the
origin and continues smoothly across the $z$-axis.  As the surface
possesses a rotational symmetry about the $z$-axis, it is self
intersecting where it crosses $\rho = 0$.  These self intersection
points are physical caustic points.  The horizon and locus clearly show the
cusp-like nature of the horizon at the caustic point. With the
parameterization described in Eq.~(\ref{fflocus}), no special
treatment is needed for handling the caustic point.

In Fig.~\ref{fig:locus} we show the initial horizons
surrounding the throats with their cusps facing each other, and the
entire surface of all photons that will ever join the horizon surface
for the full evolution to the future. This remarkable picture
completes our understanding of the location of the EH of this classic
two black hole initial data set discovered by Misner~\cite{Misner} over
30 years ago.

\section{Conclusions}

We have developed a powerful method for finding black hole event
horizons in dynamic spacetimes based on the ideas of ({\it i})
backward integration and ({\it ii}) integrating the entire null
surface.  This opens up the possibility of studying the dynamics of
event horizons in numerical relativity.

Our methods allow the determination of the horizon location to
exceptional accuracy, even in highly distorted dynamical black holes
involving strong gravitational waves, rapidly rotating black holes,
and colliding black holes.  Due to the convergence properties
of null surfaces when integrated backwards, the horizon-containing
region can be narrowed to resolutions far better than that of the
numerical simulation which created the background spacetimes.  The
width of the horizon containing region diminishes exponentially in
time. Thus, the event horizon can be located to a very small fraction of
the grid spacing.

We have shown that if the spacetime can be evolved to a point that the
black hole has returned to approximate stationarity, a
horizon-containing region can be chosen.  The precise width of this
region is unimportant, as it exponentially decreases backwards in
time.  In particular, our methods do not require the knowledge of the
apparent horizon, so they should be a useful tool for analyzing
spacetimes even in cases where the apparent horizon cannot be found
(e.g., if the time slicing does not intersect the apparent horizon).
The impact of this development on the numerical investigations of the
cosmic censorship conjecture and the hoop conjecture could prove
interesting.

Our methods allow one to locate and trace the actual generators of the
event horizon, as well as its location in spacetime.  The methods are
also able to handle caustic points on the horizon surface. The ability
to find horizons and their generators accurately and efficiently
allows one to probe geometrical and physical properties of the
horizon, including horizon oscillations, generators, membrane-paradigm
type quantities, and other previously unattainable physical
properties.  The methods for finding these quantities will be the
topic of the second paper~\cite{Masso95a} in this series.  These tools
will be important for studying the structure of event horizons in
dynamical black hole spacetimes, including colliding black holes, as
we will present in the third paper in this series\cite{Masso95b}.

\acknowledgements
We are most grateful to Kip Thorne for many discussions and for
contributing ideas to this work.  We thank Richard Isaacson and
Richard Matzner for useful discussions.  We also thank Scott Hughes,
Charles R. Keeton II, Stuart Shapiro, Saul Teukolsky, and Kevin Walsh
for allowing us to use the Cornell Photon Integration code for making
comparisons with
our code, and especially S.H., C.R.K., and K.W. who, together with one
of us (P.W.), developed some of the I/O routines used in the present
version of our code. We are grateful to Pete Anninos for help with the
two black hole studies, and to Steve Brandt for providing data for
evolved rotating black holes that were discussed in this
paper. J.M. acknowledges a Fellowship (P.F.P.I.) from Ministerio de
Educaci\'on y Ciencia of Spain. This research is supported by the
NCSA, the Pittsburgh Supercomputing Center, and NSF grants
Nos. PHY91-16682, PHY94-04788, PHY94-07882 and PHY/ASC93-18152 (arpa
supplemented).


\begin{figure}
\caption{A spherical black hole spacetime is shown in
Eddington-Finkelstein coordinates with several outgoing null rays
trajectories plotted.  The horizon is
shown as a solid line, while various photon paths are shown as dotted
lines, with their $r/M$ location marked.}
\label{fig:edfink}
\end{figure}

\begin{figure}
\caption{{\em (a)}. We show the lapse at $t=100M$ for a single
Schwarzschild black hole evolved with our 2-D axisymmetric code using
maximal slicing.  The horizon-containing region, determined by the
condition $0.1 < \alpha < 0.7$, is marked by crosses.  The outer edge
of the region is marked (o) and the inner is marked (i).  {\em
(b)}. We show the evolution of radial coordinate location $r$ of the
apparent horizon (dotted line), and the location of horizon-containing
region (solid lines) for the backward surface method applied to a
Schwarzschild spacetime evolved with our axisymmetric black hole code.
Other test surfaces are shown as dashed lines. All surfaces converge
rapidly towards the true event horizon location.  {\em (c)}. The
coordinate location of the horizon-containing region is shown as solid
lines.  A nonspherical initial trial surface is shown as a dashed
line.  The apparent horizon is shown as a dotted line.  The
nonspherical initial trial surface converges to the EH, just as do the
spherical trial surfaces.  The final horizon, at $t=0$, is shown as a
thick line. {\em (d)}. The evolution (at times $t=98.5M$, $t=98.4M$,
$t=98.3M$ and $t=98.2M$ from top to bottom, with the line AH labeling
the position of the true EH at $t=98.5M$) of a very distorted trial
surface is shown in coordinate space. The evolution breaks down,
showing that a highly distorted trial surface can
develop trouble if the parameterization of the surface is unsuitable.
See discussion in the text.}
\label{fig:ehahsph}
\end{figure}

\begin{figure}
\caption{To illustrate both the tangential drifting effect and in what sense
the EH is attractive, as discussed in the text, we show the evolution
of two photons launched at $t=100M$ right on the EH but with initial
direction not exactly normal to the EH.  They have a small but nonzero
value in the ratio of the initial angular to radial coordinate
velocities.  The trajectories of the photons are represented by dotted
lines and the times at various points on the trajectories are shown in
units of $M$. The photons drift out from the horizon and cross each other
at $t=98.0M$, producing a false horizon caustic point if these photons
were taken as horizon generators.  Tracing further backward in time,
they turn around and asymptotically approach the correct radial
location of the EH.  The radial coordinate (vertical axis) is rescaled
by the radial coordinate value of the EH, so that the EH is always at
$1$ on the vertical axis.  The horizontal axis is $\theta$ in radians.
Although the photons remain rather close to the correct EH location
throughout the trajectory, they drift substantially in the tangential
direction.  For various implications of this behavior, see discussions
in the text.}

\label{fig:tandrift}
\end{figure}

\begin{figure}
\caption{The case of a spherical black hole with a massless scalar
field falling in is shown.  The dashed line represents the apparent
horizon.   The line marked ``Event Horizon'' is obtained by
applying the backward methods to a point initially on the
apparent horizon at $t=60M$.  The dotted lines labeled ``Escaping
Photon'' and ``Trapped Photon'' represent photons integrated forward
in time from just outside and just inside the ``EH''.}
\label{fig:scalar}
\end{figure}

\begin{figure}
\caption{{\em (a)}. The area of various trial surfaces is shown for a
slightly distorted Schwarzschild spacetime.  The horizon-containing
region [between the solid lines labeled (o) and (i)] is traced
backward in time and compared to the area of the apparent horizon
(short dashed line).  Several other test surfaces are shown as long
dashed lines.  The attracting nature of the event horizon is dramatic,
as all of the trial surface integrations trace the same path after
some integration, although they start from very different initial
locations.  The insert shows an expanded view of the early time
results.  {\em All} surface integrations are shown, and are completely
indistinguishable. {\em (b)}. We show the maximum width among the
angular zones of the horizon-containing region in the slightly
distorted Schwarzschild spacetime, integrated backward in time.  We
note that the narrowing of the horizon containing region is
exponential. {\em (c)}. We compare results obtained with the backward
surface method to those obtained with the backward photon method for
the perturbed Schwarzschild spacetime.  The surface method results are
shown as solid lines, while the surfaces formed by the backward
integration of photons are shown as dashed lines.  At all times, the
results agree to within 0.05 grid zones, so that it is almost
impossible to distinguish the results on this graph.}
\label{fig:pertarea}
\end{figure}

\begin{figure}
\caption{We again show the width of the horizon-containing domain in terms
of grid zones, here for the highly distorted single black hole.  We note
that the narrowing of this region is again exponential, diminishing by a
factor of $10^5$ from $t=60M$ to $t=0$.}
\label{fig:distortcoord}
\end{figure}

\begin{figure}
\caption{We show the maximum coordinate separation between various trial
surfaces and the exact horizon in the analytic Kerr spacetime with a
rotation parameter $a/m = 0.68$.  The convergence to the exact solution is
exponential, even though for a Kerr black hole rotating this rapidly the
horizon has a very nonspherical geometry.}
\label{fig:kerr}
\end{figure}

\begin{figure}
\caption{{\em (a)}. We show the coordinate locations for several trial
surfaces for the distorted rotating black hole spacetime discussed in
the text.  The solid lines marked (i) and (o) represent the inner and
outer edges of the horizon containing region at $t=60M$.  A distorted
trial surface and another trial surface inside initially inside the
apparent horizon, shown as dashed and dotted lines respectively, were
also evolved.  All four surfaces converge quickly to the same
location. By $t=43M$ all surfaces are very close to each other and by
$t=30M$ they are practically indistinguishable.  The final location of
all surfaces, denoting the horizon location at $t=0$, is marked as a
thick line. {\em (b)}. We show the width of the horizon-containing
region in terms of grid zones.  We note that the separation exhibits
the same exponential convergence as in the non-rotating cases.}
\label{fig:kerrdistort}
\end{figure}

\begin{figure}
\caption{{\em (a)}. We plot the lapse function at $t=75M$ for the two
black hole case with $\mu = 2.2$.  The boundaries of the
horizon-containing region for the final, coalesced hole are shown as
thick lines.  The inner edge of the region, marked (i), is chosen to
be the apparent horizon, and the outer edge, marked (o), is chosen to
be at $\alpha=0.7$. {\em (b)}. We show the coordinate location of the
horizon-containing region as solid lines marked (o) and (i), and a
distorted trial surface as a dotted line.  We note that the surfaces
bifurcate when traced backwards in time (thereby coalescing forwards
in time) as expected.  The final horizon at $t=0$ is marked as a thick
line.  {\em (c)}. We show the width of the horizon-containing region
in grid zones.  Once again, we see exponential convergence.}
\label{fig:twobhcompare}
\end{figure}

\begin{figure}
\caption{The event horizon for the collision of two black holes is computed
by two
different methods.  The solid lines show the result of integrating a
trial surface backward in time (starting at the position of the apparent
horizon),
and the dashed lines show the surfaces obtained by directly
integrating the geodesic equation for a series of backward photons
starting at the same location and aimed normal to the apparent horizon
surface.  Geodesics which leave the horizon through caustics are
not displayed.  The results agree to within 0.05 grid zones throughout,
making it almost impossible to
distinguish between the solid and dashed lines.}
\label{fig:twobhphot}
\end{figure}

\begin{figure}
\caption{Here we show the evolution of the event horizon and locus of
generators of the $\mu=2.2$ two black hole collision.  Only the top
half of the system, containing one hole, is shown. The first surface
at $t=3.3M$ shows a single, elongated black hole.  The evolution of
the surface is displayed at $t=2M$, $t=1M$, ending at $t=0$.  By
$t=2M$, the surface has already crossed itself, showing two separate
holes and the locus of generators that have not yet joined the
horizon.  The cusp points where the surface intersects itself are
caustics of the horizon. }
\label{fig:twobhlocus}
\end{figure}

\begin{figure}
\caption{The coordinate location of the event horizon (thick line),
and the locus of generators (thin line) that have not yet joined the
horizon, is shown at time $t=0M$ for two black holes colliding head
on.  The $z$ coordinate line marks the symmetry axis. This locus of
generators will join the horizon through the point where the surface
intersects itself.}
\label{fig:locus}
\end{figure}

\end{document}